# An Automated Ac Susceptibility Set up Fabricated Using a Closed-Cycle Helium Refrigerator

S. Kundu and T. K. Nath[*]

*Department of Physics and Meteorology, Indian Institute of technology Kharagpur,
Kharagpur-721302, West Bengal, India.*
[*]*Email: tnath@phy.iitkgp.ernet.in*

**Abstract.** We have described here the design and operation of an automated ac susceptibility set up using a closed cycle helium refrigerator. This set up is useful for measuring linear and nonlinear magnetic susceptibilities of various magnetic materials. The working temperature range is 2 K to 300 K. The overall sensitivity of the set up is found to be $10^{-3}$ emu.

## INTRODUCTION

The measurement of ac magnetic susceptibility has been proved to be an efficient tool for characterizing a variety of magnetic materials. After the availability of commercial SQUID magnetometer, very precession measurement is now possible. Although, homemade ac susceptometers are also being used for measurement of ac susceptibility (ACS) for their certain flexibilities and low cost [1]. For example higher order or nonlinear susceptibility can be measured in a properly designed homemade susceptometer employing commercial Lock-In-Amplifier (LIA). We describe here the design and operation of a homemade ACS set up capable of measuring linear and nonlinear ACS in the temperature range 2K to 300 K.

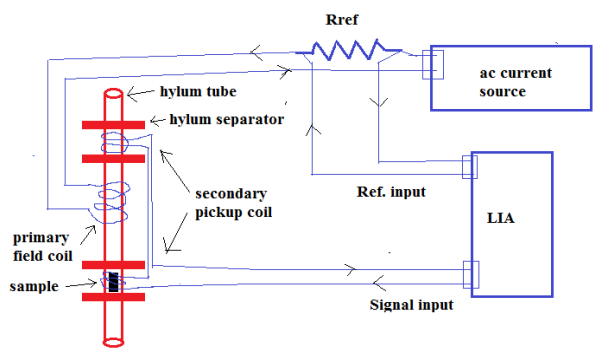

**FIGURE 1.** Schematic of the ACS coil and electrical wirings.

## BASIC PRINCIPLE

The basic electrical connection for measurement of ACS is depicted in Fig.1. A coil system comprises of a primary field coil (solenoid) and two secondary pick-up coils kept coaxially, are used to measure the ACS signal. These pick-up coils are made identical and connected in series opposition. In this case, if a sinusoidal current is passed through the primary coil the voltage induced across the pick-up coils is nearly zero. If the sample is placed in one of the pick-up coils a net voltage having same frequency as the primary current will appear which is measured employing a LIA. We have used Keithley ac current source (model-6221) for generating the ac field in the primary and Stanford research system LIA (SR 830) for voltage measurement. We used the voltage across a resistor connected in series with the primary coil as the reference voltage of the LIA. The voltage induced in the pick-up coil is proportional to the ACS of the sample. The proportionality constant has been estimated through calibration of the system by standard sample ($Gd_2O_3$). We know the magnetization of a specimen is expanded as,

$$M = M_0 + \chi_1 H + \chi_2 H^2 + \chi_3 H^3 + ...$$

Here, $\chi_1$ is the linear susceptibility and $\chi_2, \chi_3$ etc. are called the nonlinear susceptibility. By definition, ACS

$$\chi = \left|\frac{dM}{dH}\right|_{H\to 0}$$

. So it can be shown in low ac field ($h_{ac}$) limit [2,3] that the measured voltage induced in the pick-up coils at frequencies ω, 2ω, 3ω etc. will be proportional to $\chi_1$, $\chi_2$, $\chi_3$, respectively, where ω is the frequency of ac field in primary. The real and imaginary part of ACS can be measured by measuring the out of phase and in phase voltages induced in the pick-up coil with respect to the primary current. We can also superimpose a dc field by applying a dc bias current from the current source.

## DESIGN

The schematic of the coil system with basic electrical connection to the electronic equipments is shown in the Fig. 1. A hylum tube of 70 mm length and 7 mm diameter has been used as a former of the primary windings. We have used 40 SWG enameled copper wire for the primary coil. The secondary pick-up coils are wound on the primary by using a pair of annular hylum discs for each coil as shown in the Fig. 1 Roughly 3000 turns are given per coil using 45 SWG enameled copper wire.

The coil system is inserted into the variable temperature insert (VTI) of a close cycle helium refrigeration superconducting magnet system from Cryogenics Ltd., U.K. as shown in the Fig. 2. He gas is circulated in the coil zone by an oil free pump. The He pot is kept below 4 K by a separate helium compressor. Lakeshore made temperature controller (TC) 340S is employed to control the temperature of the sample. We use nonmagnetic cernox thermometer to measure the temperature. The ACS coil system is kept at a distance from the superconducting (S.C.) solenoid to avoid the residual field. The data acquisition is done automatically by employing computer and Labview software.

Some of the measured data in this ACS set up are shown in the Fig. 3 for a manganite material $Nd_{0.4}Gd_{0.3}Sr_{0.3}MnO_3$.

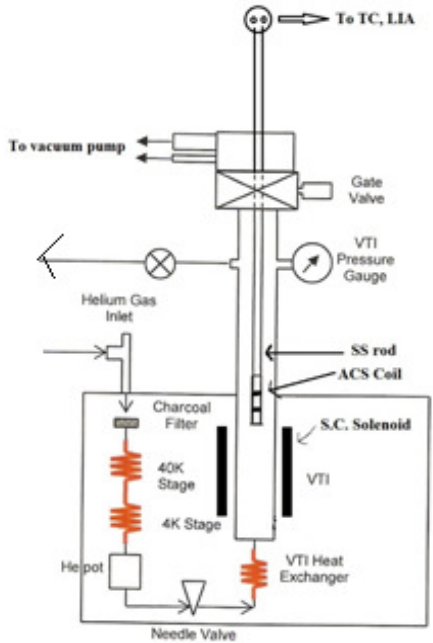

**FIGURE 2.** A schematic diagram of the set up. All the parts are labeled.

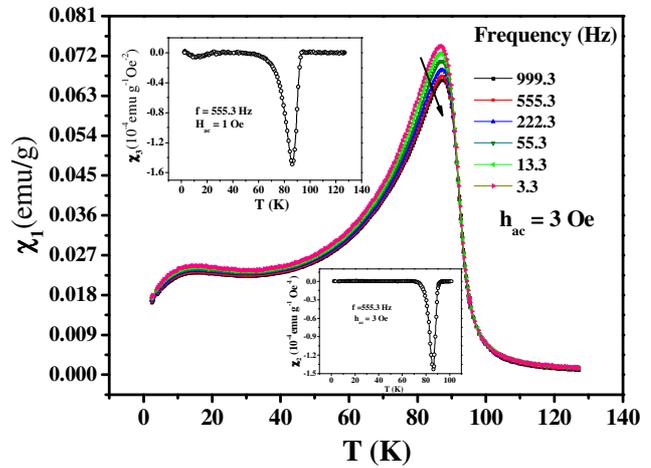

FIGURE 3. Measured linear and nonlinear susceptibilities of $Nd_{0.4}Gd_{0.3}Sr_{0.3}MnO_3$.

## REFERENCES

1. A. Bajpai and A. Banerjee, Rev. Sci. Instrum. **68**, 4075-4079 (1997).
2. T. Sato, Y. Miyako, J. Phys. Soc. Jpn. **51**, 1394-1400 (1981).
3. S. Kundu and T. K. Nath, J. Magn. Magn. Mater. **322**, 2408-2414 (2010).